\documentclass[%
 reprint,
 amsmath,amssymb,
 aps,
]{revtex4-2}
\usepackage{mleftright}
\usepackage{graphicx}
\usepackage{float}
\usepackage{dcolumn}
\usepackage{bm}
\usepackage{placeins}
\usepackage{xcolor}

\begin{document}

\preprint{APS/123-QED}

\title{Scaling invariance for the diffusion coefficient in a billiard system}

\author{$^1$Anne Kétri P. da Fonseca, $^2$Diego F. M. Oliveira, $^1$Edson D. Leonel}
\affiliation{{$^1$Departamento de Física, UNESP - Universidade Estadual Paulista, Avenida 24A, 1515, Bela Vista, Rio Claro, 13506-900, São Paulo, Brazil\\
$^2$School of Electrical Engineering and Computer Science - University of North Dakota, Grand Forks, Avenue Stop 8357, 58202, North Dakota, USA}}

\date{\today}

\begin{abstract}
We investigated the unbounded diffusion observed in a time-dependent oval-shaped billiard and its suppression owing to inelastic collisions with the boundary. The main focus is on the behavior of the diffusion coefficient, which plays a key role in describing the scaling invariance characteristic of this transition. For short times, the low-action regime is characterized by a constant diffusion coefficient, which begins to decay after a crossover iteration, thereby suppressing the unlimited growth of velocity. We demonstrate that this behavior is scaling-invariant concerning the control parameters and can be described by a homogeneous generalized function and its associated scaling laws. The critical exponents are determined both phenomenologically and analytically, including the decay exponent $\beta = -1$, previously identified in the diffusion coefficient of the dissipative standard map.

 \noindent\textbf{Keywords:} Diffusion equation, Scaling invariance, Scaling laws, Billiard systems 
\end{abstract}

\keywords{Diffusion equation, Scaling invariance, Scaling laws, Billiard systems}
\maketitle


\section{Introduction}
\label{introduction}

A diffusion process can be defined as a Markovian random process, that is, indexed by time in such a way that the future is independent of the past and satisfies a stochastic differential equation \cite{balakrishnan2008elements}. Such processes may describe the transport processes in numerous systems, such as the dissemination of pollutants in the air \cite{popp2006international}, the invasive spread of species through landscape ecology \cite{brooks2002habitat}, meteorological applications \cite{neiburger1965diffusion}, relativistic particle diffusing in an electromagnetic field \cite{relativ} and the spread of infectious diseases such as Influenza, Zika, and, more recently, COVID-19 \cite{sirs,ijapm2014influenza,zika, covid, covid2}.

An approach to investigating this diffusion is Fick's law \cite{fick1855diffusion}. Such laws describe the movement of molecules from regions of higher concentration to those of lower concentration and how this change in concentration occurs over time \cite{sethna2021statistical}. Both of these laws define a quantity $D$ that mediates the connection between these assumptions, known as the diffusion coefficient, the main observable that will be investigated in this paper. 

We focus on the diffusion coefficient for a time-dependent oval billiard, which undergoes a transition from bounded to unbounded diffusion of its root-mean-squared velocity $V_{rms}$. A billiard is a dynamical system composed of a particle or a set of non-interacting particles that experience specular collisions with a rigid boundary that confines them \cite{chernov2006chaotic}. The shape of the boundary, written in polar coordinates as a function $R_b(\theta,t)$, fundamentally influences the system's dynamics. These can be fully integrable \cite{berry1981regularity,bunimovich2005open}, fully chaotic \cite{dettmann2009survival,lozej2018aspects}, or exhibit a mixed phase space structure, as is the case for the oval billiard, which is the primary focus of this work \cite{lopac2002chaotic}.  The presence of chaos on the last two scenarios lead us to the Loskutov-Ryabov-Akinshin (LRA) Conjecture \cite{loskutov1999mechanism,loskutov2000properties}, which determines that such existence in the phase space is a sufficient, not necessary, condition for the occurrence of an unlimited energy growth, also known as Fermi Acceleration \cite{leonel2009fermi}  when a time perturbation of the boundary is introduced. The unbounded growth can be suppressed by introducing dissipation, as is the case for inelasticity in the collisions characterized by a restitution coefficient $\gamma$ \cite{leonel2016thermodynamics}. 

Although this transition from the conservative to the dissipative case is well-defined in terms of the behavior of average velocities, characterized by their scale invariance, with critical exponents related by scaling laws, such a discussion is not present regarding the diffusion coefficient. In this work, we consider a time-dependent oval billiard and the main aspects of the $V_{rms}$ behavior. We consider a set of control parameters that leads the particles to exhibit diffusive dynamics, then introduce dissipation, which leads the system to a stationary state, thereby making such diffusion finite. We also derive analytical expressions for both the velocity and the probability distribution function for the system, as well as for the diffusion coefficient $D$. The distribution mentioned above gives the probability of observing a particular particle with a given
velocity at a specific time. The expression for $D$ enables a deeper understanding of the mechanisms underlying the phase transition, while also exhibiting scaling invariance and being adequately characterized by a generalized homogeneous function. This also allows us to write scaling laws relating the critical exponents obtained for the diffusion curves. This analysis is the main result of this paper and marks our original contribution to the discussion.

This paper is organized as follows: Section \ref{xsec2} describes the model and mapping for the time-dependent oval billiard. Section \ref{xsec3} gives the solution of the diffusion equation as well as the scaling invariance observed for the diffusion coefficient. The critical exponents are also presented. In Section \ref{xsec4}, the homogeneous generalized function is obtained, as well as the analytical obtaining of the critical exponents, agreeing with the ones presented in Section \ref{xsec3}. Section \ref{xsec5} presents the final remarks and conclusions.

\section{The model and the mapping}
\label{xsec2}
The object of study of this work is the dynamics of a time-dependent oval-shaped billiard as it goes from unbounded to bounded diffusion as dissipation is introduced. The boundary is given by $R_b(\theta,t) = 1 + \epsilon[1 + \eta \cos(t)] \cos(p\theta)$, where \( \epsilon \) controls the integrability of the system where $\epsilon=0$ leads to the circular billiard, which is fully integrable, while \( \epsilon \neq 0 \) leads to a phase space containing chaos, periodic islands, and invariant spanning curves \cite{lopac2002chaotic}. The parameter \( \eta \) controls the amplitude of the temporal dependence,  \( \theta \) is the polar angle relative to the origin of the coordinate system, and \( p \) is an integer that deforms the boundary, as non-integer values would allow the escape of particles through holes in the boundary.  For a critical value $\epsilon> \epsilon_c = (1 + p^2)^{-1} $, all invariant spanning curves are destroyed \cite{oliveira2010suppressing}. The time-dependent boundary \( \eta \neq 0 \), as defined by the LRA conjecture, leads to unbounded diffusion caused by such chaotic dynamics on the static regime.

The dynamics is described by a four-dimensional nonlinear mapping $T$ for a set of dynamical variables $T(\theta_n, \alpha_n, \vert \vec{V}_{n}\vert , t_n)=(\theta_{n+1}, \alpha_{n+1}, \vert \vec{V}_{n+1}\vert , t_{n+1})$, where $\alpha$ and $\theta$ are taken about a tangent line and the origin, respectively, at the collision, $\vert \vec{V}\vert $ denotes the particle's velocity, and $t$ is the time. The coordinates for the particle are written as $X(\theta_n)=R(\theta_n,t_n)\cos(\theta_n)$ and $Y(\theta_n)=R(\theta_n,t_n)\sin(\theta_n)$. The velocity at the  $n^{th}$ collision is:
\begin{equation}
    \vec{V_n}= |\vec{V_n}|[\cos(\phi_n+\alpha_n)\hat{i} + \sin(\phi_n + \alpha_n)\hat{j}],
\end{equation}
were an auxiliary angle $\phi_n$ given by:
\begin{equation}
    \phi_n = \arctan\left[\frac{Y'(\theta_n,t_n)}{X'(\theta_n,t_n)}\right].
\end{equation}
The prime indicates derivatives with respect to $\theta_n$. The position of the particle is obtained by $R_p(t) =\sqrt{X^2 (t) + Y^2 (t)}$, and the point of collision is found for $R_p = R_b$. Specular reflection laws are written for a non-inertial reference frame of the moving boundary:
\begin{eqnarray*}
    \vec{V'}_{n+1} \cdot \vec{T}_{n+1} =  \vec{V'}_{n} \cdot \vec{T}_{n+1} \\
    \vec{V'}_{n+1} \cdot \vec{N}_{n+1} =  \vec{V'}_{n} \cdot \vec{N}_{n+1}
\end{eqnarray*}
with $\vec{T}$ and $\vec{N}$ being unit vectors for the tangential and normal components, respectively. The velocity at the $(n+1)^{th}$ collision is written as:
\begin{equation}
\vert \vec{V}_{n+1}\vert = \sqrt{(\vec{V}_{n+1}\cdot \vec{T}_{n+1})^2 + (\vec{V}_{n+1}\cdot \vec{N}_{n+1})^2}.
\end{equation}
Lastly, the updated angle $\alpha_{n+1}$ is given by
\begin{equation}
\alpha_{n+1}=\arctan\left[\frac{\vec{V}_{n+1}\cdot \vec{N}_{n+1}}{ \vec{V}_{n+1}\cdot \vec{T}_{n+1}}\right].
\end{equation}
The typical observable to analyze diffusion is given by the square root of the averaged squared velocity, defined as
\begin{equation}
 {V_{rms}}=\sqrt{\frac{1}{M} \sum^M_{i=1} \frac{1}{n}  \sum^n_{j=1} V^2_{i,j}}
\label{Eq_1}
\end{equation}
where $V_{i,j}$ indicates the velocities over each trajectory $i$ and all the $j$ different orbits. The averages are taken over $M$ different initial conditions, $n$ collisions.

\begin{figure}[h]
  \centering
    \centerline{\includegraphics[width=1\linewidth]{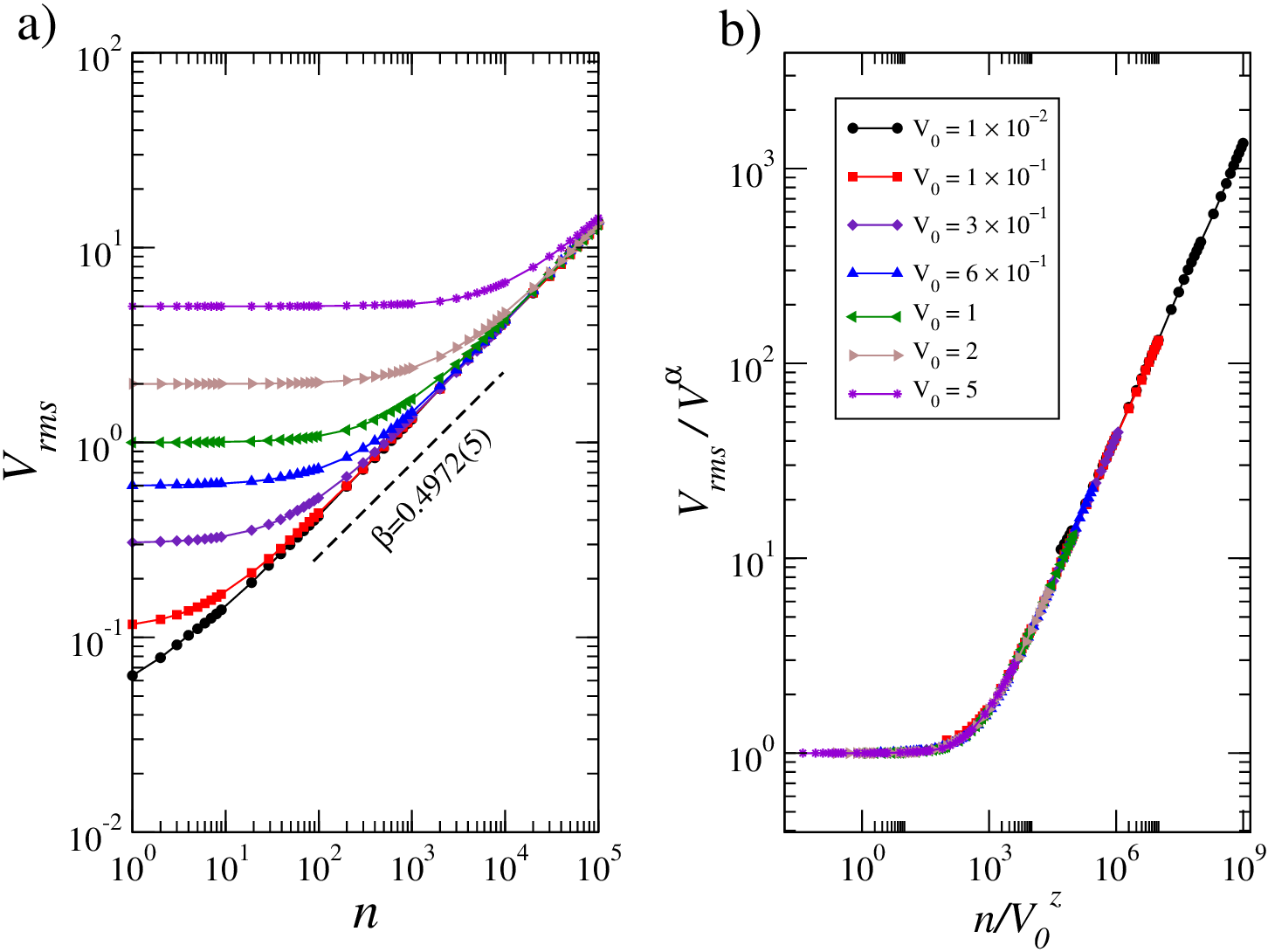}}
    \centerline{\includegraphics[width=1\linewidth]{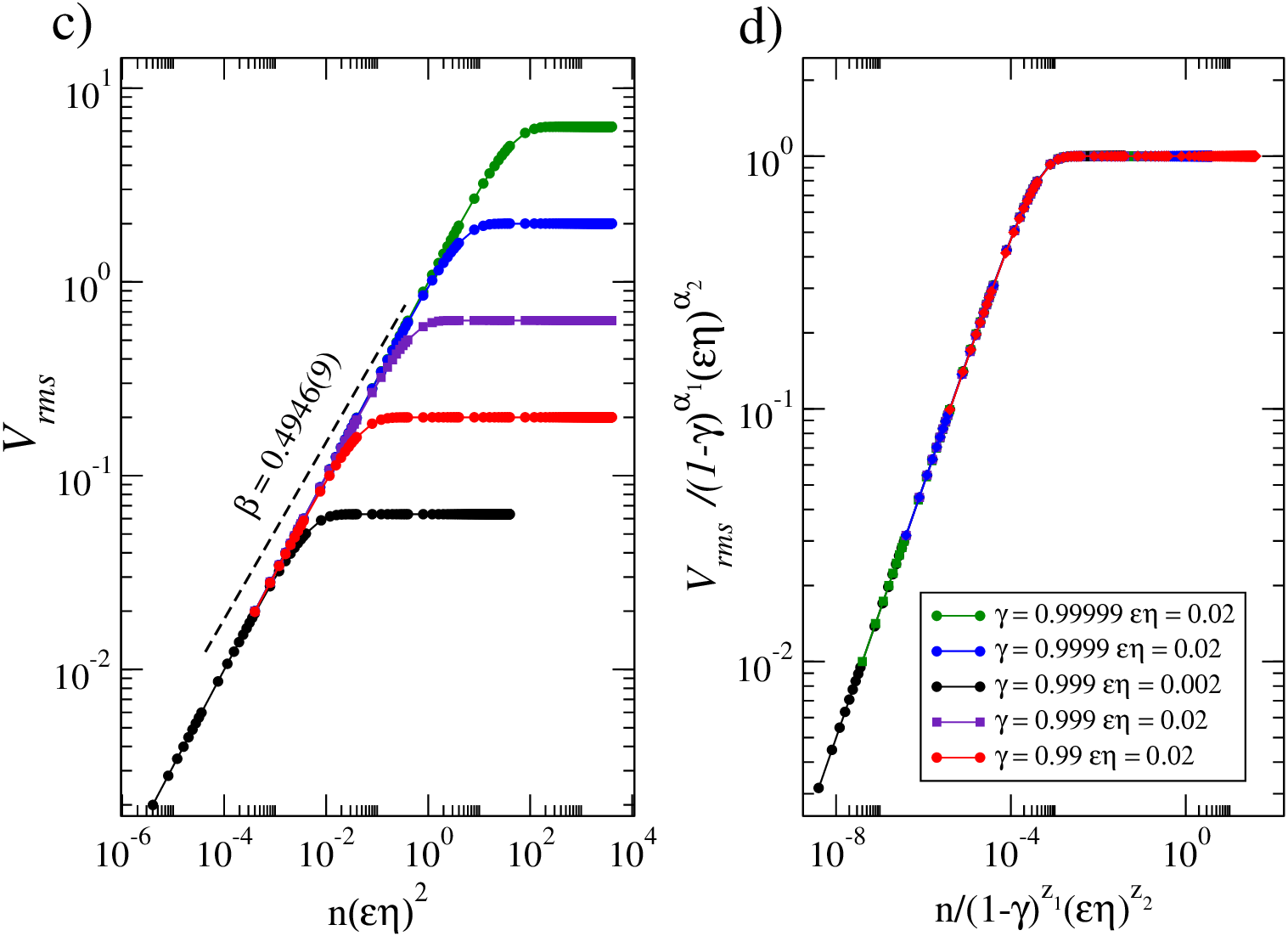}}    
  \caption{(Color online) $V_{rms}$ $vs.$ $n$ for \textit{(a)} the conservative case with  $\epsilon=0.08$, $p=3$, $\eta=0.5$ and different values of $V_0$ and \textit{(c)} the dissipative case for initial velocity $V_0 = 10^{-5}$ for different values of $\gamma$ and $\eta \epsilon$. \textit{(b)} and \textit{(d)} present the overlap of the curves in \textit{(a)} and \textit{(c)}, respectively, onto single universal curves using the adequate scaling transformations.
  }
\end{figure}

The scaling for the conservative case, as showed in Fig.1 (a), can be summarized as: (i) for sufficiently small, of the order of the velocity of the boundary, initial velocities $V_0$, $V_{rms}$ grows as a power-law described by $V_{rms} \propto n^\beta$ with $\beta \approx 0.5$; (ii) for greater values of $V_0$ an initial plateau is observed, given by $\overline{V}_{plat} \propto  V_0^\alpha$, with $\alpha\approx 1$ and (iii) the crossover from the plateau to the growth regime is written as $n_x \propto V_0^z$, with $z\approx 2$. Such behavior can also be described by a homogeneous and generalized function $ \overline{V}(n,V_0)=l\overline{V}(l^a n, l^b V_0)$, which leads to the scaling law $z=\alpha/\beta$. The transformation $V_{rms} \rightarrow V_{rms}/V^\alpha$ and $n \rightarrow n/V_0^z$ overlap all curves onto a universal plot, presented in Fig. 1 (b).

The unlimited energy growth presented for the conservative case can be suppressed considering the collisions of the particles with the boundary to be inelastic \cite{oliveira2010suppressing}. We introduce dissipation through a restitution coefficient $\gamma<1$ in the $\vec{N}$ component, where $\gamma=1$ retrieves the conservative case and the collisions are fully elastic. We rewrite the reflection laws as:
\begin{equation}
\vec{V}_{n+1} \cdot \vec{N}_{n+1} = -\gamma \vec{V}_{n} \cdot \vec{N}_{n+1} + (1 + \gamma) \vec{V_b}[t_{n+1} + Z(n)] \cdot \vec{N}_{n+1}
\end{equation}
As in the conservative case and as presented in Fig.1 (c), the behavior of the $V_{rms}$ can be summarized as: (i) for a sufficiently small number of collision \( V_{rms} \propto [(\eta \epsilon)^2 n]^\beta \), with \( \beta \approx 0.5 \); (ii) after a crossover, the velocity reaches a plateau described by \( \overline{V}_{sat} \propto (1 - \gamma)^{\alpha_1} (\eta \epsilon)^{\alpha_2} \), where \( \alpha_1 \approx -0.5 \) and \( \alpha_2 \approx 1 \) and (iii) the crossover iteration is given by \( n_x \propto (1 - \gamma)^{z_1} (\eta \epsilon)^{z_2} \), with \( z_1 \approx -1 \) and \( z_2 \approx 0 \).  Once again we write a homogeneous and generalized function $ V_{rms} [(\eta\epsilon)^2n, \eta\epsilon, (1-\gamma)]= l V_{rms}  [l^a(\eta\epsilon)^2n,l^b \eta\epsilon, l^d(1-\gamma)],$ that leads to the scaling laws $z_1=\frac{{\alpha_1}}{\beta}$ and $z_2=\frac{\alpha_2}{\beta}-2$. Once again the overlap of the curves into a single universal one is done using scaling transformations: $V_{rms} \rightarrow V_{rms}/(1-\gamma)^{\alpha_1}(\epsilon\eta)^{\alpha_2}$ and $n\rightarrow n/(1-\gamma)^{z_1}(\eta\epsilon)^{z_2}$, as presented in Fig. 1 (d).

Both results corroborate the presence of such scaling invariance for the diffusion coefficient along the transition. In Section \ref{xsec3}, we discuss how this diffusion can be obtained through the solution of a diffusion equation and what its influence is on the overall behavior of the system.

\section{The diffusion equation solution and scaling invariance}
\label{xsec3}
When considering the dynamics for the conservative case, for a sufficiently high initial velocity, the plateau observed for $V_{rms}$ occurs due to an even distribution of such velocities, resembling a Gaussian distribution, where some particles gain energy. In contrast, others lose energy due to collisions with the boundary \cite{leonel2016thermodynamics}. Such distribution progressively flattens itself until the system reaches a lower limit: the particles can't have velocities lower than the velocity of the moving boundary, given by:
\begin{equation}
   |\vec{V_b}(t_{n+1})|= \mleft.\dfrac{d{R(t)}}{d{t}}\mright\vert_{t_{n+1}} [cos(\theta_{n+1})\hat{i} + sen(\theta_{n+1})\hat{j}] \propto \eta\epsilon.
\end{equation}
This symmetry breaking leads to a scenario in which a larger fraction of the ensemble gains energy than loses it, resulting in a crossover and subsequently a diffusion in the average velocity, hence its energy \cite{oliveira2015symmetry}. Since, for such scenarios, all the regular structures in the phase space are destroyed, we assume that the spread of particles can be described using the diffusion equation.

As mentioned, Fick's law can be used to describe the flow of a current of particles $J$ crossing a line in an interval of time. Such a flow is proportional to the gradient of concentration as $J=-D\frac{\partial P(V,n)}{\partial V}$, where the sign comes from the current running opposite to the potential gradient. Here, $P(V,n)$ denotes the probability of observing a particle with a velocity $V$ at an instant of time $n$. The conservation of the number of particles, guaranteed by integer values of $p$, allows us to write
\begin{equation}
    \frac{\partial P(V,n)}{\partial n} + \frac{\partial J}{\partial V}=0
\end{equation}
and thus
\begin{equation}
    \frac{\partial P(V,n)}{\partial n}=\frac{\partial D}{\partial V}\frac{\partial P}{\partial V} + D\frac{\partial^2 D}{\partial V^2}.
\end{equation}
Here, $D$ is the diffusion coefficient, indicating how fast particles disperse through the phase space due to random motion. We assume that the dependence of $D$  on the variation from $n$ to $n+1$ is slow and smooth, leading the first term on the right side of equation (9) to vanish. 

It's noticeable from this result that obtaining the analytical expression of $P(V,n)$ allows us to retrieve all the main observables of the system. To do so, we impose as boundary conditions: (i) $P(V,n)|_{V\rightarrow0} =  P(V,n)|_{V\rightarrow\infty} = 0$ limiting the unbounded growth of diffusion showed in Fig. 1 (c); (ii) $P(V, 0) = \delta(V-V_0)$, an initial condition that ensures that at $n=0$ all particles are starting from the same initial velocity distributed in $M$ different initial conditions uniformly distributed over $\alpha$, $\theta$ and $t$. The solution to equation (9) comes from the image method formalism combined with the distribution for a semi-infinite line \cite{balakrishnan2008elements,pre2025} as:
\begin{equation}
    P(V,n)= \frac{\tau}{\sqrt{4\pi D n}} \left[ e^{ \left( \frac{-(V-V_0)^2}{4Dn} \right)} - e^{\left( \frac{-(V+V_0)^2}{4Dn} \right)} \right],
\end{equation}
where $\tau=erf\left(V_0/\sqrt{4Dn}\right)$ emerges from the normalization. 

\begin{figure}[H]
    \centering
    \includegraphics[width=\linewidth]{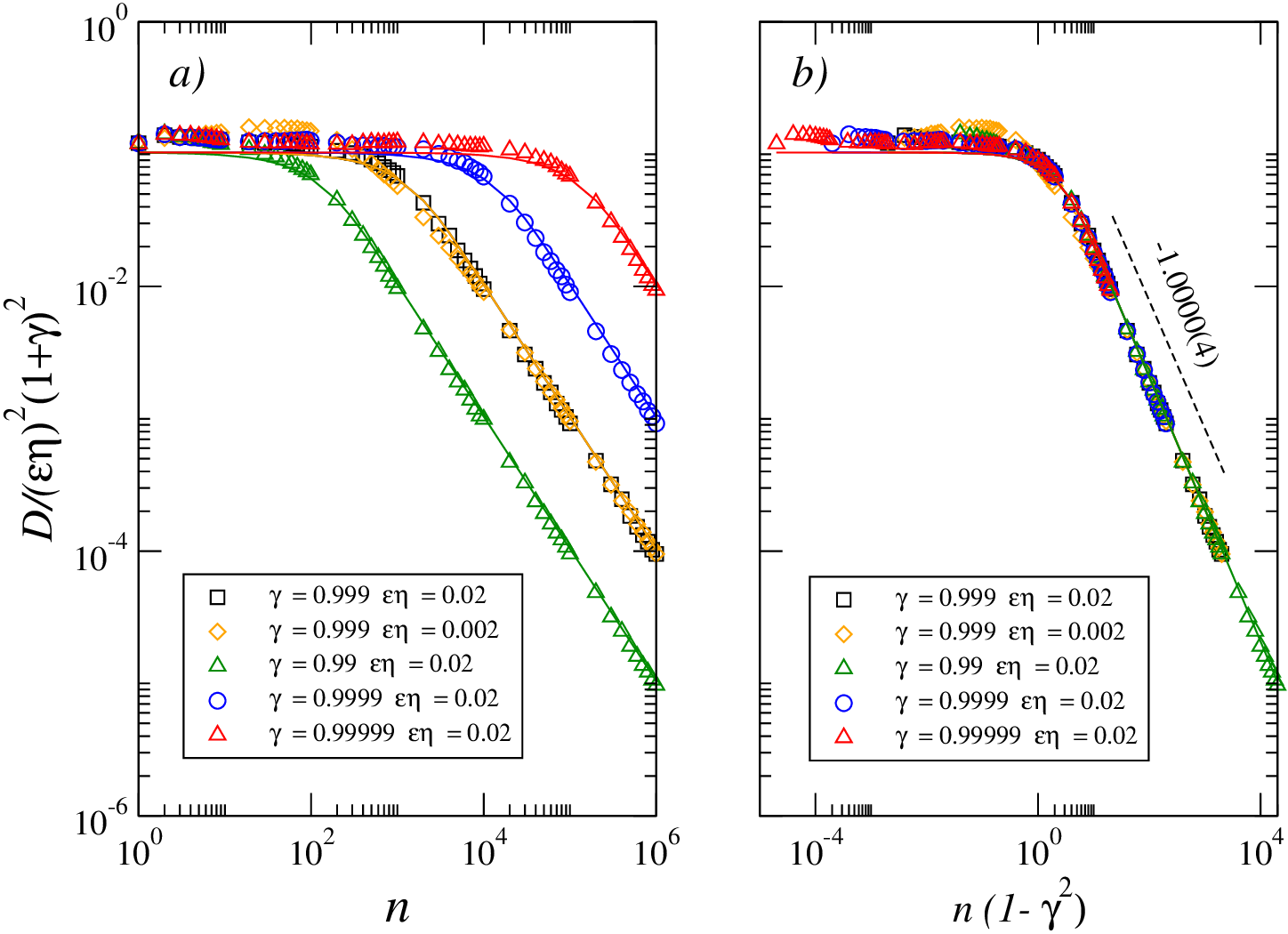}
    \caption{Plot of curves of $D/(\epsilon\eta)^2(1+\gamma)^2$ (a) $vs. \ n$ for different control parameters and (b) $vs. \ n(1-\gamma^2)$ overlapping the curves in (a) onto a single universal plot. In both cases, the symbols represent numerical curves while the continuous lines are obtained from equation (17). }
    \label{fig:enter-label}
\end{figure}

We can now obtain the behavior of the diffusion coefficient $D$ as the mean squared displacement $D=\frac{\overline{V^2_{n+1}}-\overline{V^2_{n}}}{2}$. We recur to $ \vert \vec{V}_{n+1}\vert$ and perform the quadratic velocity average for the intervals $\theta \in [0,2\pi]$, $\alpha \in [0,\pi]$ and $t\in [0,2\pi]$ \cite{leonel2016thermodynamics}.
\begin{equation}
     \overline{V^2_{n+1}} = \frac{(\gamma^2+1)\overline{V^2_{n}}}{2} + \frac{(1+\gamma)^2\eta^2\epsilon^2}{8}.
\end{equation}
Substituting this result in the expression for $D$, we obtain:
\begin{equation}
    D=\frac{(\gamma^2-1)\overline{V^2_{n}}}{4} + \frac{(1+\gamma)^2\eta^2\epsilon^2}{16}.
\end{equation}
Now it's necessary to obtain $\overline{V^2_{n}}$. We assume that, for a large ensemble of particles, $
    \overline{V^2_{n+1}}-\overline{V^2_{n}}=\frac{\overline{V^2_{n+1}}-\overline{V^2_{n}}}{(n+1)-n} \simeq \frac{d\overline{V^2}}{dn}.$ The integration of this result gives 
    \begin{equation}
\overline{V^2}(n)=\overline{V_0^2}e^{\frac{(\gamma^2-1)n}{2}} + \frac{(1+\gamma)}{4(1-\gamma)}\eta^2\epsilon^2[1-e^{\frac{(\gamma^2-1)n}{2}}].
\end{equation}
Performing the average of the above equation, so that it is comparable to the quantity $V_{rms}$, and making the necessary simplifications, the final solution is:
\begin{gather}
    \langle \overline{V^2}(n)\rangle = \frac{1}{n+1}\left(\overline{V_0^2} - \frac{(1+\gamma)}{4(1-\gamma)}\eta^2\epsilon^2\right)\left[\frac{1-e^{\frac{(\gamma^2-1)(n+1)}{2}}}{1-e^{\frac{(\gamma^2-1)}{2}}}\right] \nonumber \\
    + \frac{(1+\gamma)}{4(1-\gamma)}\eta^2\epsilon^2 .
\end{gather}
From this expression we can retrieve the limits for small and sufficiently large $n$. The first is obtained for $n=0$, where $\langle \overline{V^2}(0)\rangle=V_0$. As for $n\rightarrow \infty$ we obtain
\begin{equation}
    V_{sat}=\sqrt{\frac{(1+\gamma)}{(1-\gamma)}{}}\frac{\eta\epsilon}{2}
\end{equation}
which retrieves the critical exponents $\alpha_1=-0.5$ and $\alpha_2=1$. The power law growth can also be retrieved with a Taylor Series expansion of equation (14) for $\gamma \approx1$:
\begin{equation}
    V_{rms} \simeq \frac{\eta\epsilon\sqrt{n}}{2}.
\end{equation}
From the scaling laws, we obtain $\beta =0.5$, confirming the critical exponents $z_1=-1$ and $z_2=0$.

Substituting equation (14) in the expression of $D$ and making the necessary simplifications, we obtain
\begin{gather}
    D = \frac{1}{n+1}\left(\overline{V_0^2} - \frac{(1+\gamma)}{4(1-\gamma)}\eta^2\epsilon^2\right)\times \nonumber\\\left[\frac{1-e^{\frac{(\gamma^2-1)(n+1)}{2}}}{1-e^{\frac{(\gamma^2-1)}{2}}}\right]\frac{(\gamma^2-1)}{4}.
 \end{gather}
Now analyzing the limits of equation (17): (i) for $n=0$ $ D(n) \propto (1+\gamma)^2\eta^2\epsilon^2/16$ when $V_0 \simeq 0$; (ii) $n\rightarrow\infty$ leads to $D(n)=0$, so that the particle is no longer under diffusion and the system reached the stationary state presented in Fig. 1 (c) and (d).

Figure 2 presents plots of $D(n)$ for different control parameters, where the symbols represent the results obtained numerically from the mean squared displacement, while the continuous lines are obtained from equation (17). The curves remain constant for a short time, associated with the growth of the $V_{rms}$ curves presented in Fig. 1(c). After a crossover, $D(n)$ experiences a decay in power law with exponent $-1$, as presented in Figure 2 (b), eventually reaching zero. This crossover is the same as that experienced for the $V_{rms}$, from growth to saturation, mentioned in the previous section.

The Taylor Series expansion of equation (14) also gives us
\begin{equation}
    D \approx \frac{(1+\gamma)^2(\eta\epsilon)^2}{16}\left(1 - \frac{(1-\gamma^2)n}{4}+ \ ...\right)
\end{equation}
and estimating, for the crossover from the initial plateau to the decay regime $1-(1-\gamma^2)n_x/4 \simeq 0$, leading to
\begin{equation}
    n_x \simeq (1-\gamma^2)^{-1}.
\end{equation}
This result can also be confirmed by obtaining the crossover interaction $n_x$ for a set of $D(n) \ \ vs. n$ curves and analyzing its dependence on $(1-\gamma^2)$. The plot for $n_x \ \ vs. (1-\gamma)^2$ is presented in Figure 3 (a), for which the power law fitting retrieves the exponent $-1$ found in equation (19). Such an equation may also be rewritten as  $n_x  \simeq (1-\gamma)^{-1}$, once only this term provides the criticality as we approach the transition in $\gamma\rightarrow1$. The plot for $n_x \ \ vs. (1-\gamma)$ presented in Figure 3 (b) confirms this behavior, where it is useful to define $z_2$ as the exponent obtained from the power law fitting, as will be shown in the next section.

Thus, a suitable scaling variables are $D\rightarrow D/(\epsilon\eta)^2(1+\gamma)^2$ and $n \rightarrow n(1-\gamma^2)$. Figure 2 (b) shows the overlap of the $D(n)$ curves after such scaling transformations.  
\FloatBarrier

\section{Homogeneous function and analytical solution}
\label{xsec4}
We now present an analytical procedure to obtain all the critical exponents and write the adequate scaling laws for the diffusion coefficient. It is useful to start by analyzing the asymptotic behaviors of equation (17). As mentioned, at the initial iteration $n=0$, $ D(n) \propto (1+\gamma)^2\eta^2\epsilon^2/16$ when $V_0 \simeq 0$. As for sufficiently long times, that is $n\rightarrow\infty$, $D(n)=0$. Let us now discuss the behavior during decay. 

As mentioned, $\gamma$ is a value very close to but smaller than 1. Therefore, it is useful to define $\gamma=1-\delta$ where $\delta \ll1$. Such substitution wields
\begin{equation}
    D = \frac{1}{n+1}\left(\overline{V_0^2} - \frac{(2-\delta)}{4\delta}\eta^2\epsilon^2\right)\left[\frac{1-e^{-\delta(n+1)}}{1-e^{-\delta}}\right]\frac{-2\delta}{4}.
\end{equation}
Is possible to rewrite $e^{-\delta} \approx 1- \delta$ and $1-e^{-\delta}\approx 1-1-\delta =\delta$. Furthermore, we know that, as $n$ grows, the exponential term goes to zero substantially faster than the $1/(n+1)$ term, which allows us to write equation (20) as:
\begin{equation}
     D = \frac{1}{n+1}\left(\overline{V_0^2} - \frac{(2-\delta)}{4\delta}\eta^2\epsilon^2\right)\frac{-1}{2} \simeq  \frac{\eta^2\epsilon^2}{4(1-\gamma)(n+1)}.
\end{equation}
Therefore, we not only found an expression for the decay but also confirmed that such behavior is dependent on the control parameter as $D(n)\propto \eta^2\epsilon^2(1-\gamma)^{-1}n^{-1}$. Finally, the crossover iteration number is obtained by equating the results of the saturation and decay behaviors described above, isolating $n$. We then have
\begin{equation}
n_x \propto (1-\gamma)^{-1}(1+\gamma)^{-2}
\end{equation}
It is worth noting that this same quantity can be rewritten knowing that $(1-\gamma)(1+\gamma)=(1-\gamma^2)$, leading to $n_x \propto(1-\gamma^2)^{-1}$, as seen in the previous section.

\begin{figure}[h]
\centering
    \centerline{(a)\includegraphics[width=0.9\linewidth]{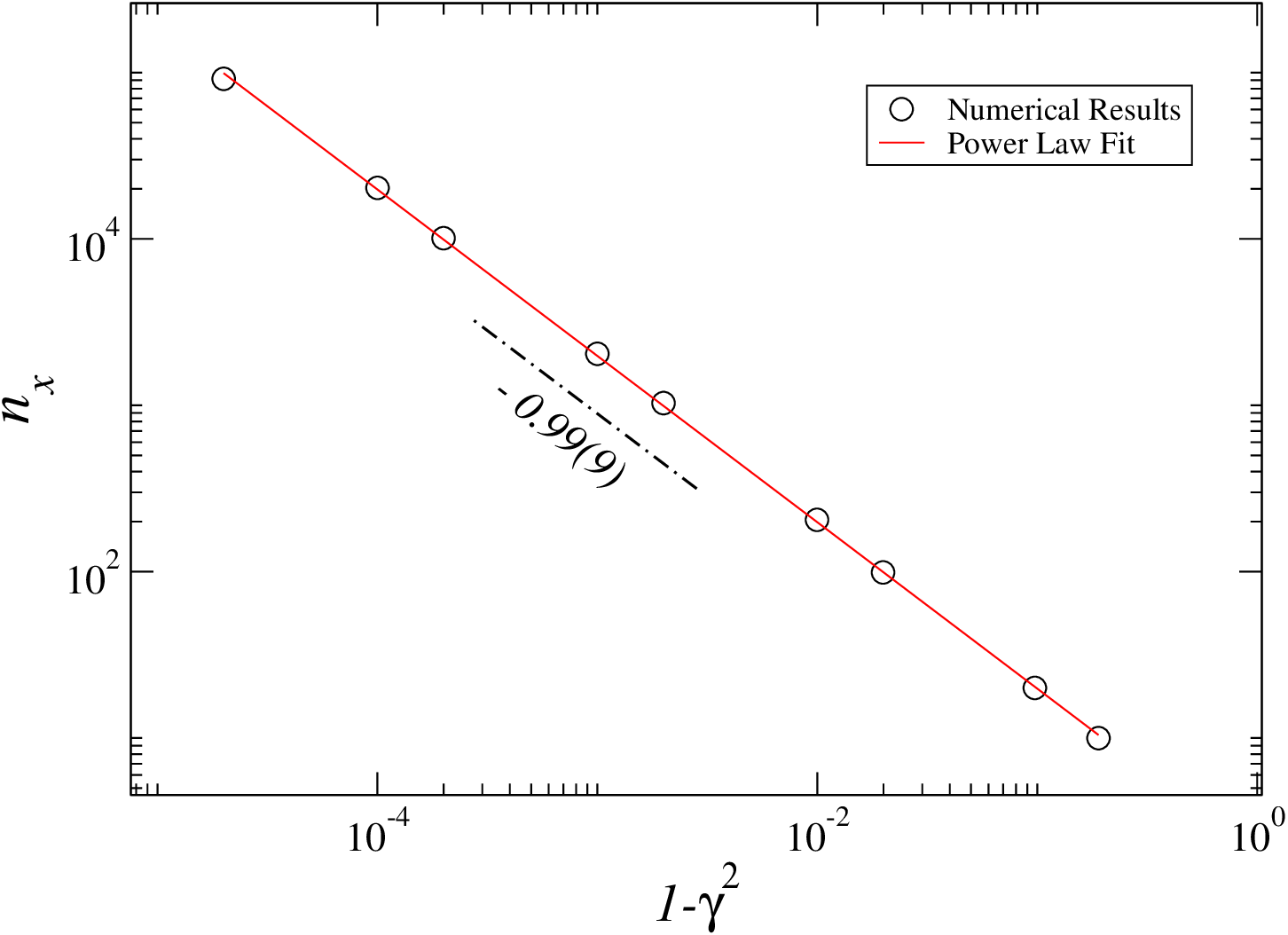}}
    \centerline{(b)\includegraphics[width=0.9\linewidth]{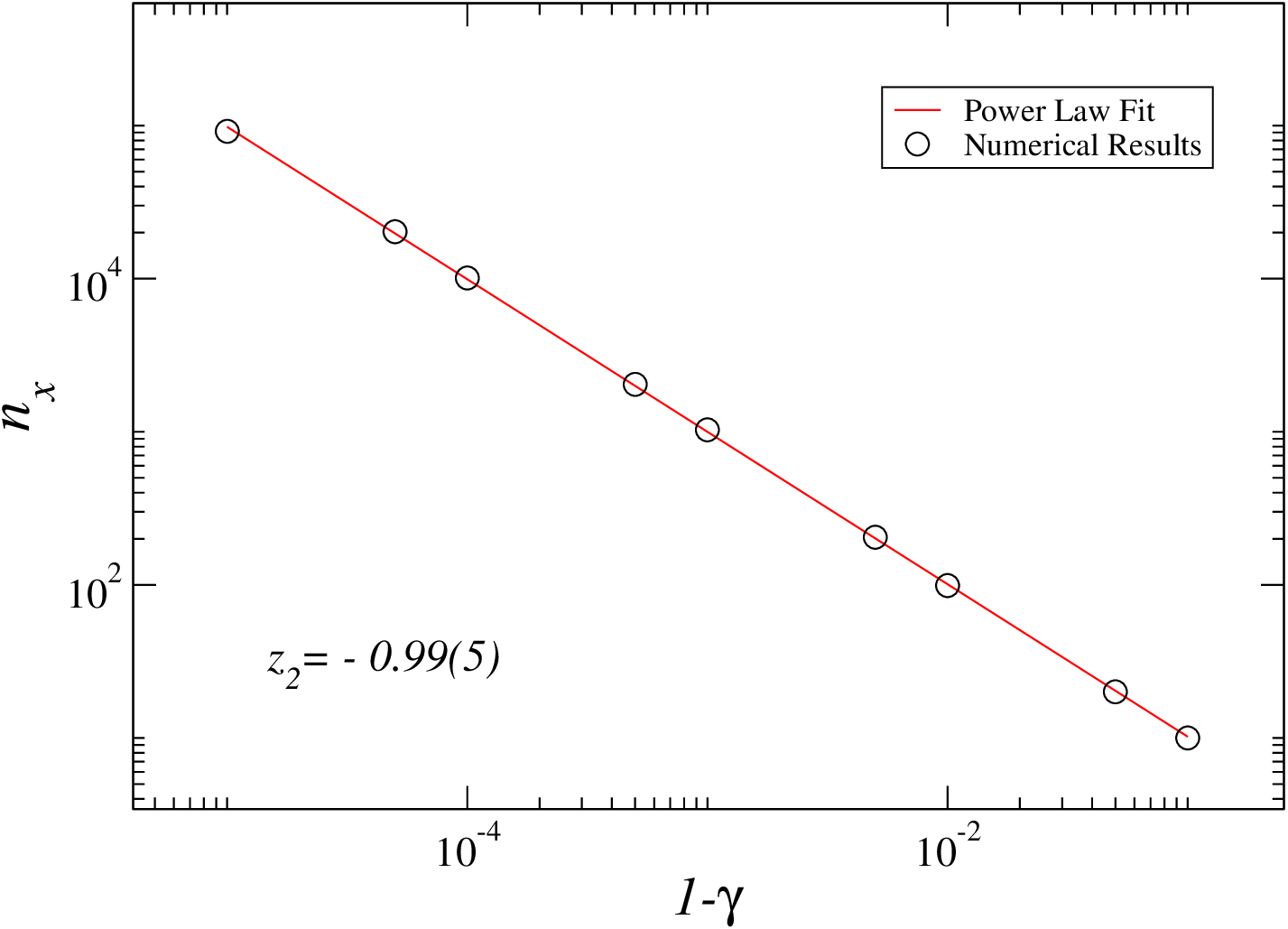}}    
\caption{(a) Plot of $D(n)$ $vs.$  $(1-\gamma^2)$. A power law fitting yielded an exponent of $-0.99(9)$. (b) Plot of $D(n)$ $vs.$  $(1-\gamma)$ with a power law fit giving $z_2=-0.99(5)$.}
\label{expoentes}
\end{figure}

Now, analyzing the behavior of Figure \ref{fig:enter-label}, we can propose the following scale hypotheses: (i) For values of $n \ll n_x$, the saturation regime is described by $D_{sat} \propto (1+\gamma)^{\alpha_1}(\eta^2\epsilon^2)^{\alpha_2}$, where $\alpha_1$ and $\alpha_2$ are the saturation exponents; (ii) For $n \gg n_x$, the decay is described by $D(n) \propto \eta^2\epsilon^2[(1-\gamma)n]^\beta$ where $\beta$ is the decay exponent; (iii) The \textit{crossover} iteration is given by $n_x \propto (1+\gamma)^{z_1}(1-\gamma)^{z_2} $.

Thus, the three hypotheses allow the behavior of the diffusion coefficient to be described by a single generalized homogeneous function given by:
\begin{gather}
D \ \left[(1-\gamma)n,\eta^2\epsilon^2,(1+ \gamma)\right]=\nonumber\\l \ D \ \left[l^a(1-\gamma)n, l^b\eta^2\epsilon^2,l^d(1+ \gamma)\right]
\end{gather}
where $l$ is a scale factor to which we can assign certain values, as we did previously for $\overline{V}$, so that some argument of the expression for $D$ is equal to $1$. We begin by taking $l^a(1-\gamma)n=1 \rightarrow l = \left[(1-\gamma)n\right]^{-1/a}$, thus:
\begin{gather}
      D \left[(1-\gamma)n,\eta^2\epsilon^2,(1+ \gamma)\right]=  \nonumber \\
      \left[(1-\gamma)n\right]^{-1/a} D \ [1,l^{-b/a}\eta^2\epsilon^2, l^{-d/a}(1+\gamma)]
 \end{gather}
 where we assume that $D \ [1,l^{-b/a}\eta^2\epsilon^2, l^{-d/a}(1+\gamma)]$ exhibits constant behavior for sufficiently long times. Comparing this case with the hypothesis describing power law decay, we have $ \left[(1-\gamma)n\right]^{-1/a} =[(1-\gamma)n]^\beta$ and $\beta =-1/a$. Now for $l^b\eta^2\epsilon^2=1 \rightarrow l=(\eta^2\epsilon^2)^{-1/b}$ we repeat the procedure, writing:
 \begin{gather}
       D \left[(1-\gamma)n,\eta^2\epsilon^2,(1+ \gamma)\right]=  \nonumber \\(\eta^2\epsilon^2)^{-1/b} D \left[l^{-a/b}(1-\gamma) n,1,l^{-d/b}(1+ \gamma)\right]
 \end{gather}
 where we assume that $D \left[l^{-a/b}(1-\gamma) n,1,l^{-d/b}(1+ \gamma)\right]$ exhibits constant behavior for short times. Comparing this case with the hypothesis describing saturation, we have $ (\eta^2\epsilon^2)^{-1/b} =(1+\gamma)^{\alpha_1}(\eta^2\epsilon^2)^{\alpha_2}$ and $\alpha_2 =-1/b$. Lastly, repeating the procedure for $l^d(1+\gamma) =  1 \rightarrow l = (1+\gamma)^{-1/d}$
 \begin{gather}
      D \left[(1-\gamma)n,\eta^2\epsilon^2,(1+ \gamma)\right]= \nonumber \\(1+\gamma)^{-1/d} D \  \left[l^{-a/d}(1-\gamma)n,l^{-b/d}\eta^2\epsilon^2,1\right]
 \end{gather}
 once again assuming $D \ \left[l^{-a/d}(1-\gamma)n,l^{-b/d}\eta^2\epsilon^2,1\right]$ exhibits constant behavior for short times. Comparing this case with the hypothesis that describes saturation, we have $(1+\gamma)^{-1/d} =(1+\gamma)^{\alpha_1}(\eta^2\epsilon^2)^{\alpha_2}$ and $\alpha_1 =-1/d$.

 Now, to correlate the exponents, we assume that in the crossover iteration $n_x$, regimes (i) and (ii) can be equated
\begin{equation}
    (1+\gamma)^{\alpha_1}(\eta^2\epsilon^2)^{\alpha_2}=\eta^2\epsilon^2[(1-\gamma) n_x]^\beta 
\end{equation}
simplifying the terms $(\eta^2\epsilon^2)$ and directly substituting the expression for $n_x$ into equation (39) we obtain:
\begin{gather}
      (1+\gamma)^{\alpha_1}(\eta^2\epsilon^2)^{\alpha_2-1}=[(1-\gamma) (1+\gamma)^{z_1}(1-\gamma)^{z_2} ]^\beta \nonumber \\=(1+\gamma)^{z_1\beta }(1-\gamma)^{\beta(1+z_2)}  
 \end{gather}
A direct comparison between the $(\eta^2\epsilon^2)$ exponents leads to $\alpha_2-1=0$ and $\alpha_2=1$. We also recover $(z_2+1)\beta=0$, which leads to $z_2=-1$ for $\beta \neq 0$ and $z_1=\alpha_1/\beta$. This result for $z_2$ is confirmed by numerical results, as shown in Figure 3(b).

We can also recover such exponents by comparing limits appropriate to each case and the scale hypotheses. Comparing the results obtained using expression (21) with those obtained graphically, we find $\beta=-1$. This result leads to $z_1=\alpha_1/\beta=-\alpha_1$ and $z_1=-2$. A comparison with equation (22) also leads to $n_x= (1+\gamma)^{z_1}(1-\gamma)^{z_2}=(1+\gamma)^{-2}(1-\gamma)^{-1} $, again leading to $z_1=-1$ and $z_2=-2$.  Finally, the scale hypothesis for the saturation, when compared with the expression for the plateau, retrieves $\alpha_2=1$ and $\alpha_1=2$.

Thus, all critical exponents necessary for characterizing the system were adequately recovered, revealing how the relationship between diffusion and each of the control parameters, $\epsilon$, $\eta$, and $\gamma$, occurs. The first two are primarily responsible for the diffusion length
and the particle's elementary excitation. As for $\gamma$, it limits and bounds the diffusion throughout the phase space. In addition, it was possible to express analytically and phenomenologically the variation of $D(n)$ over time, which is slow from instant $n$ to $(n+1)$. This approximation, which considers the constant coefficient for solving the diffusion equation, enabled us to characterize the main observables of the system.
\section{Summary and conclusions}
\label{xsec5}

We characterize the behavior of the diffusion coefficient for the transition from bounded to unbounded diffusion in a time-dependent oval billiard.  In the static case, the phase space exhibits a mixed structure. The chaotic sea in the static scenario leads to average energy growth for a time-dependent boundary. The introduction of dissipation by inelastic collisions limits this unbounded energy diffusion, leading to a constant plateau for the $V_{rms}$, with a phase transition being observed as the restitution coefficient $\gamma\rightarrow1$.

The probability density $P(V,n)$ to observe a particle with specific action at collision $n$ was obtained by solving the diffusion equation. We assume that such diffusion varies slowly with time and obtain an analytical expression for $D(n)$ from the mean squared displacement. For short times, the regime of low action is described by a
constant diffusion coefficient that starts decaying after a crossover iteration, suppressing unlimited
velocity growth and thus the unbounded diffusion. We prove that this behavior is scale-invariant to the control parameters $\eta$, $ \epsilon$, and $\gamma$.  An initial estimative for the crossover iteration leads to $n_x \simeq (1-\gamma^2)^{-1} \simeq (1-\gamma)^{-1}$, with the exponent $-1$ signaling the same universality class of the dissipative standard mapping \cite{scaling2025}. A single universal plot confirms that the suitable scaling transformations are $D\rightarrow D/(\epsilon\eta)^2(1+\gamma)^2$ and $n\rightarrow n(1-\gamma^2)$.

Finally, we analyze the behavior of $D(n)$ for each of the limit situations: the initial plateau, the crossover and the decay with power law. With such analytical expressions, we propose a set of scaling hypotheses and a generalized homogeneous equation that describes the system. Such a procedure allowed us to successfully recover all the scaling exponents necessary to explain the variation of the behavior of the diffusion coefficient with respect to the control parameters. This result elegantly demonstrates the union between established tools of statistical mechanics and a well-established chaotic system, assuming that once all regular structures in the phase space are destroyed, the spread of particles can be described using the diffusion equation. The scale invariance of this exponent, and its agreement with the already known behavior of the $V_{rms}$ curves with dissipation, after the transition, are also strong indicators of the validity of this approximation, allowing further insight into the mechanisms behind the changes in particle diffusion and the associated phase transition.
\section*{Acknowledgements}
A.K.P.F. acknowledges CAPES (No. $88887.990665/2024-00$) for financial support. E.D.L. acknowledges support from Brazilian agencies CNPq (No. $301318/2019-0, 304398/2023-3)$ and FAPESP (No. $2019/14038-6$ and No. $2021/09519-5)$.

\bibliography{main}

\end{document}